\documentclass[12pt]{article}

\usepackage{amssymb}
\usepackage{amsmath}
\usepackage{amscd}
\usepackage{latexsym}
\usepackage{graphicx}
\usepackage{url}

\usepackage{cite}
\usepackage{caption}

\newcommand{\be}{\begin{equation}}
\newcommand{\ee}{\end{equation}}
\newcommand{\Dlt}{\Delta}

\newcommand{\bt}{\beta}
\newcommand{\vp}{\varphi}
\newcommand{\ep}{\varepsilon}
\newcommand{\al}{\alpha}
\newcommand{\ra}{\rightarrow}

\newcommand{\Gm}{\Gamma}

\newcommand{\lbd}{\lambda}

\begin{document}

\begin{center}

{\Large{\bf Self-similar interpolation in high-energy physics} \\ [5mm]

V.I. Yukalov$^{*}$ and S. Gluzman } \\ [3mm]

{\it
Bogolubov Laboratory of Theoretical Physics, \\
Joint Institute for Nuclear Research, Dubna 141980, Russia }

\end{center}

\vskip 5cm

\begin{abstract}

A method is suggested for interpolating between small-variable and
large-variable asymptotic expansions. The method is based on
self-similar approximation theory resulting in self-similar root
approximants. The latter are more general than the two-sided Pad\'{e}
approximants and modified Pad\'{e} approximants, including these as
particular cases. Being more general, the self-similar root approximants
guarantee the accuracy that is not worse, and often better, than that
of the Pad\'{e} approximants. The advantage of the root approximants
is in their unambiguous definition and in the possibility of their
construction, even when Pad\'{e} approximants cannot be defined.
Conditions for the unique definition of the root approximants are
formulated. Several examples from high-energy physics illustrate the
method.

\end{abstract}

\vskip 1cm

{\parindent = 0pt
{\bf PACS numbers}: 02.30.Lt, 02.30.Mv, 02.60.Gf, 03.65.Db, 05.10.Cc,
                    11.10.Jj, 11.15.Tk, 11.25.Pm

\vskip 2cm

{\bf $^*$corresponding author}: V.I. Yukalov

{\bf E-mail}: yukalov@theor.jinr.ru   }

\newpage

\section{Introduction}

A very often met problem in high-energy physics is the necessity of
constructing an analytical expression uniformly describing a function
$f(x)$ in the whole interval of its domain, say, in $[0,\infty]$, when
only asymptotic expansions are known for the small-variable limit
$x \ra 0$ and the large-variable limit $x \ra \infty$. The variable $x$
can represent, e.g., a coupling constant.

The standard way of treating such problems is the use of the so-called
two-sided, or two-point, or multipoint Pad\'{e} approximants \cite{Baker_1}.
In many cases, these approximants provide a reasonable interpolation
between the small-variable and large-variable limits. However, this method
has some weak points, as discussed in Refs. \cite{Baker_1,Baker_2,Gluzman_3}.
One of them is the occurrence of defects, such as spurious zeroes and poles.
The other problem is the ambiguity in choosing one of the approximant
variants from the Pad\'{e} table. The important limitation is the requirement
for the compatibility of expansions in the small-variable and large-variable
limits. The latter means the following. The standard situation in many
problems is when, in the small-variable limit, one has an expansion in
integer powers, $x^n$, while the large-variable expansion exhibits the
behavior $x^\beta$, with a noninteger power $\beta$. Since the large-variable
behavior of a Pad\'{e} approximant $P_{M/N}$ is $x^{M-N}$, this implies
that the integer power $M-N$ is not compatible with the noninteger $\beta$.
To overcome the problem of incompatibility, Baker and Gammel \cite{Baker_4}
suggested to use the fractional powers of Pad\'{e} approximants
$P_{M/N}^\gamma$, choosing the power $\gamma$ so that $(M-N) \gamma = \beta$.
The simplest case of the Baker-Gammel method is the polynomial approximant
$P_{M/0}^\gamma$ in a fractional power $\gamma = \beta/M$. The Baker-Gammel
method allows one to correctly represent the leading term of the large-variable
behavior, although the subleading terms not always can be uniquely defined
\cite{Sen_5,Honda_6}.

In the present paper, we suggest an original method of interpolation between
small-variable and large-variable expansions. This method allows one to
construct an analytical expression uniformly approximating the sought function
in the whole domain $[0,\infty]$ and reproducing both the small-variable and
large-variable expansions. The uniqueness conditions are formulated allowing
for a unique definition of all parameters. The method is more general than
that of standard Pad\'{e} approximants as well as the Baker-Gammel method of
fractional Pad\'{e} approximants. Therefore, by construction, the accuracy
of our method is not worse, and often better, than that of the latter methods,
with the advantage of being uniquely defined. The approach is illustrated by
several examples from high-energy physics.

\section{Method of self-similar interpolation}

Let us be interested in a physical quantity represented by a real function
$f(x)$ of a real variable $x \in [0,\infty]$. However, the explicit form of
this function is not known, since it is defined by complicated equations
allowing only for deriving asymptotic expansions in the vicinity of two ends,
where $x \ra 0$ and $x \ra \infty$.

For instance, in the small-variable limit, we have
\be
\label{1}
 f(x) \simeq f_k(x) \qquad ( x \ra 0 ) \;  ,
\ee
with the series
\be
\label{2}
 f_k(x) = f_0(x) \left ( 1 + \sum_{n=1}^k a_n x^n \right ) \; ,
\ee
where $f_0(x)$ is a known function. In many cases, the latter enjoys the form
\be
\label{3}
 f_0(x) = A x^\al \qquad ( A \neq 0 ) \;  ,
\ee
with $\alpha$ being any real number.

And in the large-variable limit, we can get
\be
\label{4}
 f(x) \simeq f^{(p)}(x) \qquad ( x \ra\infty) \;  ,
\ee
with the series
\be
\label{5}
  f^{(p)}(x) = \sum_{n=1}^p b_n x^{\bt_n} \; ,
\ee
where the powers $\beta_n$ are real numbers arranged in the descending order
$$
 \bt_{n+1} < \bt_n \qquad ( n = 1,2,\ldots,p-1) \; .
$$
The values of the powers $\beta_n$ can be any, either integer or fractional.

In what follows, it is convenient to deal with the reduced function $f(x)/f_0(x)$,
normalized so that in the small-variable limit,
$$
 \frac{f(x)}{f_0(x)} \simeq \frac{f_k(x)}{f_0(x)} \qquad ( x \ra 0) \; ,
$$
we would have the simple asymptotic form
$$
 \frac{f_k(x)}{f_0(x)}  = 1 + \sum_{n=1}^k a_n x^n \;.
$$

The interpolation problem consists in constructing such a representation
for the sought function $f(x)$ that would reproduce the small-variable, as
well as large-variable expansions (2) and (5), providing an accurate
approximation for the whole domain $[0,\infty]$.

It is worth stressing that this formulation of interpolation problem is
rather general. In the majority of cases, practically all realistic problems
can be reduced to this representation employing a change of variables. Also,
the small-variable and large-variable limits are conditional, since it is
always possible to interchange them by introducing the variable $t=1/x$, or
more generally, $t=1/x^\mu$, with a positive $\mu$.

The method, we suggest here, is based on self-similar approximation theory
\cite{Yukalov_7,Yukalov_8,Yukalov_9,Yukalov_10,Yukalov_11} combining the
ideas of optimized perturbation theory, optimal control theory, dynamical
theory, and renormalization-group approach. The main ideas of the theory
are as follows. The transfer from an approximate form $f_k$ to the form
$f_{k+1}$ is represented as a motion of a dynamical system in discrete
time, whose role is played by the approximation order $k = 0,1,2\ldots$.
The evolution equation of the dynamical system represents a kind of
self-similar relation, where the name {\it self-similar approximation theory}
comes from. The trajectory of the dynamical system, by construction, is
bijective to the sequence $\{f_k\}$. The dynamical system in discrete time,
that is, a cascade, can be embedded into a dynamical system in continuous
time, that is, into a flow. The convergence of a sequence $\{f_k\}$ to its
effective limit is equivalent to the convergence of the flow trajectory to
a fixed point, which, in this way, corresponds to the sought function $f$.
The motion is governed by control functions guaranteeing fast convergence
to the fixed point. The stability of the method is characterized by the
map multipliers of the related dynamical system. We shall not repeat here
all this machinery that has been expounded in all mathematical details in
Refs. \cite{Yukalov_7,Yukalov_8,Yukalov_9,Yukalov_10,Yukalov_11} and
thoroughly described in review articles \cite{Yukalov_12,Yukalov_13}, but
we shall use the results of this approach.

Using the self-similar approximation theory for the purpose of interpolation
between two asymptotic expansions, we come
\cite{Yukalov_13,Yukalov_14,Gluzman_15,Yukalov_16} to the {\it self-similar
root approximant}
\be
\label{6}
 \frac{f_k^*(x)}{f_0(x)} = \left ( \left ( \ldots ( 1 + A_1 x )^{n_1}
+ A_2 x^2 \right )^{n_2} + \ldots + A_k x^k \right )^{n_k} \; .
\ee

First of all, we see that setting here all powers $n_j = \pm 1$, we can
obtain different Pad\'{e} approximants. And if the powers $n_j = \pm 1$,
except the leading $n_k$ that is found from the leading term of the
large-variable expansion, then we get the modified Pad\'{e} approximants
of Baker and Gammel \cite{Baker_4}. However, such a choice of the powers
$n_j$ is too restrictive and arbitrary. The parameters of approximants,
to be uniquely defined, have to be prescribed by the available small-variable
and large-variable expansions.

In our previous publications, the root approximants (6) were used so that
the powers $n_i$ and parameters $A_i$ were defined through the one-sided
expansion, say, the large-variable expansion, while the other expansion, e.g.,
small-variable expansion, was not reproduced. The attempts to find all values
of $n_i$ and $A_i$ from the small-variable expansion resulted in the equations
with multiple solutions. Such a nonunique definition of the parameters is, of
course, unsatisfactory. Now, we aim at generalizing the use of the root
approximants (6) in such a way that would allow us to uniquely define all
powers $n_i$ and parameters $A_i$ and that both the small-variable as well
as the large-variable expansions be reproduced.

We may notice that the root approximant (6) can be identically rewritten as
\be
\label{7}
\frac{f_k^*(x)}{f_0(x)} = A_k^{n_k} x^{kn_k} \left ( 1 +
\frac{B_{k-1}}{x^{m_{k-1}}} \left ( 1 + \frac{B_{k_2}}{x^{m_{k-2}}}
\ldots \frac{B_1}{x^{m_1}} \left ( 1 + \frac{1}{A_1x} \right )^{n_1}
\right )^{n_2} \ldots \right )^{n_k} \; ,
\ee
with the parameters
\be
\label{8}
B_j = \frac{A_j^{n_j}}{A_{j+1}} \qquad ( j = 1,2,\ldots,k-1)
\ee
and powers
\be
\label{9}
 m_j = j+1 - jn_j \qquad ( j = 1,2, \ldots, k-1 ) \;  .
\ee

On the other hand, the large-variable expansion (5) can be represented in
the form
$$
 f^{(p)}(x) = b_1 x^{\bt_1} \left ( 1 + \frac{b_2}{b_1} \;
x^{\bt_2-\bt_1}
\left ( 1 + \frac{b_3}{b_2}\; x^{\bt_3-\bt_2} \ldots
\right. \right.
$$
\be
\label{10}
\ldots
\left. \left.
\frac{b_{p-1}}{b_{p-2}}\; x^{\bt_{p-1}-\bt_{p-2}} \left (
1+ \frac{b_p}{b_{p-1}} \; x^{\bt_p - \bt_{p-1}} \right ) \right )
\ldots  \right ) \;  .
\ee
Expanding the above expressions (7) and (10) in powers of $1/x$ and
equating the similar terms, we find that these expansions are uniquely
defined provided that Eq. (3) is valid,
\be
\label{11}
 A_k^{n_k} = \frac{b_1}{A} \qquad ( k = p ) \;  ,
\ee
the largest power $n_k$ is given by the relation
\be
\label{12}
 k n_k = \bt_1 -\al \qquad (\al\neq \bt_1 ) \;  ,
\ee
and the other powers satisfy the equations
\be
\label{13}
 m_j = \bt_{k-j} - \bt_{k-j+1} \qquad ( j = 1,2,\ldots,k-1) \;  .
\ee
In this way, all powers $n_j$ of the root approximant (6) can be uniquely
defined through the {\it uniqueness conditions}
\be
\label{14}
 j n_j = j+1 - \bt_{k-j} + \bt_{k-j+1} \qquad  ( j = 1,2,\ldots,k-1) \; .
\ee

It may happen that the number of terms in the small-variable and
large-variable asymptotic expansions are not the same, $k \neq p$. Sometimes,
just a single term of the large-variable expansion is known ($p=1)$, while
several terms of the small-variable expansion are available. How then the
uniqueness conditions (14) will be changed?

Let us assume that just the leading term of the large-variable behavior is
known:
\be
\label{15}
 f(x) \simeq B x^\bt \qquad ( x \ra \infty) \;  .
\ee
It is easy to notice that the general expansion (5) is reducible to the
asymptotic form (15) by setting $\beta_n = \beta$. Then the uniqueness
condition (14) reduces to the equality
\be
\label{16}
n_j = \frac{j+1}{j} \qquad ( k > p = 1 ) \;   ,
\ee
where $j = 1,2,\ldots,k-1$, while the leading power reads as
\be
\label{17}
 n_k = \frac{\bt-\al}{k} \qquad ( k = 1,2,\ldots) \;  .
\ee
All parameters $A_i$ are uniquely defined from the small-variable expansion.

In the general case, we have $k$ terms of the small-variable expansion and
$p$ terms of the large-variable expansion. Therefore, to satisfy these
expansions, the root approximant must be of order $k+p$, possessing $k+p$
parameters $A_j$, among which $k$ parameters $A_j$ are defined by the
accuracy-through-order procedure from the small-variable expansion and the
remaining $p$ parameters are defined from the large-variable expansion. But
then we also need to have $k+p$ equations for determining $k+p$ powers $n_j$,
while only $p$ terms of the large-variable expansion are given. How all
powers $n_j$ could be found in such a case?

Fortunately, large-variable expansions practically always enjoy the following
nice property. The difference
\be
\label{18}
\Dlt\bt_j \equiv \bt_j - \bt_{j+1} \qquad ( j = 1,2,\ldots,k-1)
\ee
between the nearest-neighbor powers is invariant:
\be
\label{19}
 \Dlt\bt_j = \Dlt\bt = const \qquad ( j = 1,2,\ldots,k-1) \;  .
\ee
In that case, with the leading power being always given by an equation of
type (12) and with all remaining powers $n_j$ being defined by the
uniqueness condition (14), we now have
\be
\label{20}
 n_{k+p} = \frac{\bt_1-\al}{k+p} \; , \qquad
j n_j = j+1 - \Dlt\bt \qquad ( j = 1,2,\ldots,k+p-1) \;  .
\ee
Then both the small-variable as well as the large-variable expansions can be
satisfied, uniquely defining all parameters $A_j$, with $j=1,2,\ldots,k+p$.

Below, we illustrate the method by several examples from high-energy physics.

\section{Supersymmetric Yang-Mills circular Wilson loop}

The $\mathcal{N} = 4$ supersymmetric Yang-Mills theory, in the limit of large
number of colors $N$ and strong t'Hooft coupling $\lambda = g^2 N$ is taken
sometimes as a model for hot QCD \cite{Blaizot_17}. Since there exists an
exactly calculable expression for the $SU(N)$ circular Wilson loop
\cite{Erickson_18,Drukker_19}, it is useful to start with this case, for which
the accuracy of approximations can be explicitly estimated.

The exact circular Wilson loop is given by
\be
\label{21}
 W(\lbd) =  \frac{2}{\sqrt{\lbd}} \; I_1(\sqrt{\lbd}) \;  ,
\ee
where $I_1$ is a modified Bessel function of the first kind. In the
weak-coupling limit, one has
\be
\label{22}
W(\lbd) \simeq e^{\sqrt{\lbd}} \left ( 1 - \lbd^{1/2} + \frac{5}{8} \; \lbd -
\; \frac{7}{24} \; \lbd^{3/2} + \frac{7}{64} \; \lbd^2 \right ) \qquad
(\lbd \ra 0 ) \;   ,
\ee
and in the strong coupling limit,
\be
\label{23}
 W(\lbd) \simeq e^{\sqrt{\lbd}} \left ( \sqrt{ \frac{2}{\pi} } \; \lbd^{-3/4} - \;
\frac{3}{4\sqrt{2\pi} } \; \lbd^{-5/4} \right ) \qquad
(\lbd \ra \infty ) \;  .
\ee

Introducing the change of the variables as
\be
\label{24}
f(x) \equiv W(\lbd(x) ) \; , \qquad \lbd = x^2 \;   ,
\ee
we obtain the weak-coupling limit
\be
\label{25}
f(x) \simeq e^x \left ( 1 - x + \frac{5}{8}\; x^2 - \; \frac{7}{24} \; x^3 +
\frac{7}{64} \; x^4 \right ) \qquad (x \ra 0 )
\ee
and the strong-coupling limit
\be
\label{26}
 f(x) \simeq e^x \left ( \sqrt{ \frac{2}{\pi} } \; x^{-3/2} - \;
\frac{3}{4\sqrt{2\pi} } \; x^{-5/2}  \right ) \qquad (x \ra \infty ) \;   .
\ee

According to rule (20), we find
\be
\label{27}
 n_1 = n_2 = n_3 = n_4 = n_5 = 1 \; , \qquad n_6 = -\; \frac{1}{4} \;  .
\ee
Then the corresponding root approximant becomes
\be
\label{28}
 f_6^*(x) = e^x \left ( 1 + A_1 x + A_2 x^2 + A_3 x^3 + A_4 x^4
+ A_5 x^5 + A_6 x^6 \right )^{-1/4} \;  ,
\ee
acquiring the form of a modified Pad\'{e} approximant, with
$$
A_1 = 4 \; , \qquad A_2 = \frac{15}{2} \; , \qquad A_3 = \frac{26}{3} \; ,
$$
$$
A_4 =\frac{653}{96} \; , \qquad A_5 = \frac{3\pi^2}{8} \; ,
\qquad A_6 = \frac{\pi^2}{4} \;   .
$$
The maximal deviation of the root approximant $f_6^*$ from the function,
corresponding to the exact Wilson loop $W$, is $0.003$. This accuracy is
the same as that of the Pad\'{e} approximant $P_{5/7}$, studied in
Ref. \cite{Banks_20} and requiring the knowledge of twice more 
terms of the asymptotic expansions.

As has been mentioned above, small-variable and large-variable expansions
can be interchanged by a change of variables. Thus, in the present case,
we may use the change
\be
\label{29}
\vp(t) \equiv W(\lbd(t) ) \; , \qquad \lbd = \frac{1}{t^2} \;   .
\ee
Then, dealing with the variable $t$ and following the general rule,
we get the root approximant
\be
\label{30}
 \vp_6^*(t) = \sqrt{\frac{2}{\pi} } \; e^{1/t} t^{3/2} \left ( 1 + B_1 t
+ B_2 t^2 + B_3 t^3 + B_4 t^4 + B_5 t^5 + B_6 t^6 \right )^{-1/4} \; ,
\ee
with the parameters
$$
B_1 = \frac{3}{2} \; , \qquad B_2 = \frac{653}{24\pi^2} \; , \qquad
B_3 = \frac{104}{3\pi^2} \; ,
$$
$$
B_4 = \frac{30}{\pi^2} \;  \qquad B_5 = \frac{16}{\pi^2} \; ,
\qquad B_6 = \frac{4}{\pi^2} \;   .
$$
Approximants (28) and (30) coincide with each other.

This example is also interesting demonstrating how the modified Pad\'e
approximants naturally arise in our method. That is, it is shown that our
method includes the modified Pad\'{e} approximants as a particular case.
Such a reduction, of course, happens not always, but rather rarely. The
root approximants (6) enjoy a more general form than Pad\'{e} approximants,
because of which they can provide better accuracy.

\section{Planar cusp anomalous dimension in supersymmetric Yang-Mills theory}

In the $\mathcal{N} = 4$ supersymmetric Yang-Mills theory, in the limit of
large angle, the planar cusp anomalous dimension is linear in angle, with a
coefficient $\Gamma (g)$ that is the cusp anomalous dimension of a light-like
Wilson loop, which depends only on the coupling $g$. The weak-coupling and
strong-coupling expansions
\cite{Korchemsky_21,Gubser_22,Frolov_23,Kotikov_24,Beisert_25,Correa_26} are
\be
\label{31}
\Gm(g) \simeq 4g^2 -\; \frac{4\pi^2}{3}\; g^4 + \frac{44\pi^4}{45}\; g^6
- 8 \left [ \frac{73\pi^2}{630} + 4\zeta^2(3) \right ] g^8
\qquad (g \ra 0 )
\ee
and, respectively,
\be
\label{32}
\Gm(g) \simeq 2g - \; \frac{3\ln 2}{2\pi} \qquad (g \ra \infty) \;  .
\ee

The corresponding root approximant reads as
\be
\label{33}
 \Gm_5^*(g) = 4g^2 \left ( \left ( \left ( \left ( \left (  1 +
A_1 g^2 \right ) ^{3/2} + A_2 g^4 \right )^{5/4} + A_3 g^6 \right )^{7/6}
+ A_4 g^8 \right )^{9/8} + A_5 g^{10} \right )^{-1/10} \;  ,
\ee
where
$$
A_1 = \frac{256\pi^2}{189} \; , \qquad A_2 = \frac{13376\pi^4}{59535} \; ,
\qquad
A_3 = \frac{32}{6751269} \left [ 54091 \pi^6 + 12859560\zeta^2(3) \right ] \; ,
$$
$$
 A_4 = 256 \left ( \frac{15\ln 2}{\pi}\right )^{8/9} \; , \qquad
A_5 = 1024 \; .
$$
This form (33) practically coincides with the Pad\'{e} approximant $P_{5/6}$
given in Ref. \cite{Banks_20}, differing from it only by $1 \%$.

\section{Spinor mass in heterotic string theory}

The $SO(32)$ spinor mass in heterotic string theory admits \cite{Sen_5,Banks_20}
a perturbative weak-coupling expansion
\be
\label{34}
M(g) \simeq g^{1/4} \left ( 1 + 0.23 g^2 \right ) \qquad ( g \ra 0 )
\ee
and a dual expansion in the limit of strong coupling,
\be
\label{35}
M(g) \simeq g^{3/4} \left ( 1 + 0.351 g^{-1} \right ) \qquad ( g \ra \infty ) \; .
\ee

The root approximant for this case is
\be
\label{36}
M_3^*(g) = g^{1/4} \left ( \left ( \left ( 1 + A_1 g^2 \right )^{3/2} +
A_2 g^4 \right )^{5/4} + A_3 g^6 \right )^{1/12} \;   ,
\ee
with
$$
 A_1 = 1.472 \; , \qquad A_2 = 3.159299 \; , \qquad
A_3 = 1 \;  .
$$
This form is very close to the Pad\'{e} approximant $P_{4/1}$ calculated
in Ref. \cite{Sen_5}, the maximal difference being only $0.5 \%$.

\section{Ground-state energy for Schwinger model}

The Schwinger model \cite{Schwinger_27,Banks_28} is a lattice gauge theory
in $(1+1)$ dimensions representing Euclidean quantum electrodynamics with
a Dirac fermion field. It possesses many properties in common with QCD,
such as confinement, chiral symmetry breaking, and charge shielding. For
this reason, it has become a standard test bed for the study of numerical
techniques.

Here we consider the ground state of the model, corresponding to a vector
boson of mass $M(x)$ as a function of the variable $x = m/g$, where $m$
is electron mass and $g$ is the coupling parameter having the dimension of
mass, so that $x$ is dimensionless. The energy is given by the relation
$E = M - 2m$. The small-$x$ expansion for the ground-state energy
\cite{Carrol_29,Vary_30,Adam_31,Striganesh_32} is
\be
\label{37}
 E(x) \simeq 0.5642 - 0.219 x + 0.1907 x^2 \qquad ( x \ra 0 ) \; .
\ee
In the large-$x$ limit \cite{Striganesh_32,Coleman_33,Hamer_34,Hamer_35},
we have
\be
\label{38}
 E(x) \simeq 0.6418 x^{-1/3} - \; \frac{1}{\pi} \; x^{-1} -
0.25208 x^{-5/3} \qquad ( x \ra \infty ) \; .
\ee

The corresponding root approximant is
\be
\label{39}
 E_5^*(x) = A \left ( \left ( \left ( \left ( \left ( 1 + A_1 x \right )^{4/3}
+ A_2 x^2 \right )^{7/6} + A_3 x^3 \right )^{10/9} + A_4 x^4 \right )^{13/12} +
A_5 x^5 \right )^{-1/15} \; ,
\ee
where
$$
A = 0.5642 \; , \qquad A_1 = 3.109547 \; , \qquad A_2 = 3.640565 \; ,
$$
$$
A _3= 4.028571 \; , \qquad A_4 = 1.070477 \; , \qquad A_5 = 0.144711 \;  .
$$

The accuracy of this approximant $E_5^*$ can be compared to data obtained
in other calculations, density matrix renormalization group, $E_{DMRG}$,
by Byrnes et al. \cite{Byrnes_36} and fast moving frame estimates, $E_{FMFE}$
by Kr\"{o}ger and Scheu \cite{Kroger_37}. This energy has also been calculated
by using variational perturbation theory \cite{Byrnes_38}, which however has
been found to be rather complicated and having no advantage over other
techniques, such as Pad\'{e} approximants \cite{Hamer_35}. Adam \cite{Adam_39}
used renorm-ordered perturbation theory, although his results are less accurate
than $E_{DMRG}$ and $E_{FMFE}$. In Table 1, we compare the latter data with
our result $E_5^*$. Also, we present the energy $E_{PA}$ found by means of
the Pad\'{e} approximant $P_{5/6}(x^{1/3})$. All results are close to each
other, being practically the same in the frame of calculational errors.

\section{Prediction of large-variable expansions}

The great advantage of the method of self-similar root approximants, as
compared to Pad\'{e} approximants, is that root approximants can predict
the correct behavior of sought functions at large variables, being based on
small-variable expansions. Such a prediction is principally impossible by
means of Pad\'{e} approximants, when the small-variable and large-variable
expansions contain incompatible powers \cite{Baker_1,Baker_2,Gluzman_3},
for example, when the small-variable expansion is in integer powers, as in
Eq. (2), while the large-variable expansion, as in Eq. (5), is in fractional
powers.

To illustrate this basic advantage, let us consider the ground-state energy
for the Schwinger model, studied in the previous section. Suppose, only the
small-variable expansion (37) is available and the powers $-1/3$ and $-1$ of
expansion (38) are known. But no coefficients from the large-variable expansion
are given. With the two terms of expansion (37), we can construct the root
approximant
\be
\label{40}
 E_2^*(x) = A \left (  ( 1 + A_1 x )^{4/3} + A_2 x^2
\right )^{-1/6} \;  ,
\ee
in which all parameters are found from the small-variable expansion (37):
$$
 A = 0.5642 \; , \qquad A_1 = 1.746721 \; , \qquad
A_2 = 0.458024 \;  .
$$
By extending this expression to large variables, we get
\be
\label{41}
 E_2^*(x) \simeq 0.642616 x^{-1/3} \qquad ( x \ra \infty) \;  .
\ee
As is seen, the value $0.6426$ very well approximates the first coefficient
$0.6418$ of the exact expansion (38).

Moreover, taking into account this prediction (41), we can construct the root
approximant
\be
\label{42}
 E_3^*(x) = A \left ( \left (  ( 1 + A_1 x )^{4/3} + A_2 x^2
\right )^{7/6} + A_3 x^3 \right )^{-1/9} \;  ,
\ee
where again the parameters $A$, $A_1$, and $A_2$ are defined by the
small-variable expansion (37), resulting in
$$
 A = 0.5642 \; , \qquad A_1 = 2.245784 \; , \qquad
A_2 = 1.336080 \; , \qquad A_3 = 0.309979 \;  .
$$
At large variables, Eq. (42) yields
\be
\label{43}
 E_3^*(x) \simeq 0.642616 x^{-1/3} - 0.322985 x^{-1}
\qquad ( x \ra \infty) \;   .
\ee
Both coefficients here are close to those of the exact expansion (38).
The value of the second coefficient $0.322985$ has to be compared
with $1/\pi = 0.318310$.

\section{Remarks on scheme and scale invariance}

It is worth noting that self-similar approximation theory 
\cite{Yukalov_7,Yukalov_8,Yukalov_9,Yukalov_10,Yukalov_11}, used for 
deriving self-similar root approximants, considered in the present paper, 
employs the ideas of renormalization group in a sense that is different 
from this notion in quantum field theory, although being close 
mathematically. In self-similar approximation theory, the transfer from 
one approximation of order $k$ to another, say, $k+1$, is considered as 
a motion in discrete time $k$. Then the sequence of approximations can be 
treated as a cascade. Embedding the cascade into a flow makes it possible 
to pass from discrete time $k$ to continuous time $t$. The flow evolution 
is represented by the renormalization-group equation of the type 
$df(x)/dt = v(x)$, where $v(x)$ is a flow velocity. Solving the latter 
equation iteratively leads to self-similar approximants for the sought 
function $f(x)$.    

In quantum field theory, such as QCD or QED, there is the known problem
of scheme and scale dependence of truncated series. The sought function
$f(x)$ is not scheme-scale dependent, while its truncated series, as in
Eq. (2), generally, can be dependent on both. Thus, if $f(x)$ is an
observable quantity that is a function of the coupling parameter
$x = \alpha_s/\pi$, then both the coefficients $a_n$, as well as the
coupling parameter $x$, can depend on renormalization scheme and scale.
The dependence of $x$ on scale comes from the renormalization-group equation
$dx/dt = \beta(x)$, where $t$ is a scale shift and $\beta(x)$ is a
Gell-Mann-Low function. It has been shown that Pad\'{e} approximants
reduce the scale dependence, when the Gell-Mann-Low function is taken
in one-loop approximation \cite{Gardi_40,Brodsky_41}, and a generalized
approach \cite{Cvetic_42,Cvetic_43,Cvetic_44,Cvetic_45} has been developed
for achieving perturbative scale invariance for any number of loops.

In the cases we have considered, the situation is different from the
mentioned quantum-field {\it extrapolation} problems, since we investigate
the {\it interpolation} method, where both the small and large variable
limits of the sought function are given. In our case, there can arise the
question of invariance with respect to a parameter, if expansion (2) is not
unique, which is analogous to the scheme dependence. For example, there
can exist different perturbative schemes, associated with different values
of a parameter $t$, so that the coefficients $a_n = a_n(t)$ in expansion (2)
depend on this parameter, while the sought function $f(x)$ should not depend
on such parameters. Then the truncated series $f_k(x,t)$ also depend on
the parameter, as a result of which the root approximant $f^*_k(x,t)$
includes the dependence on $t$. However, in the interpolation problem, the
small-variable asymptotic form $f_0(x)$ is given, being not dependent on
auxiliary parameters, as well as the large-variable limit is also assumed
to be available, say, as the asymptotic behavior (15), where $B$ does not
depend on $t$. This implies that the small-variable limit
\be
\label{44}
 \lim_{x\ra 0} \; \frac{f_k^*(x,t)}{f(x)} = \lim_{x\ra 0} \;
  \frac{f_k^*(x,t)}{f_0(x)} = 1
\ee
is fixed, together with the large-variable limit
\be
\label{45}
 \lim_{x\ra \infty} \; \frac{f_k^*(x,t)}{f(x)} = \lim_{x\ra \infty} \;
  \frac{f_k^*(x,t)}{Bx^\bt} = 1 \; .
\ee
For instance, if the small-variable behavior is given by Eq. (3) and only
the large-variable asymptotic form (15) is available, then the root
approximant is
\be
\label{46}
f_k^*(x,t) = A x^\al \left ( \left ( \ldots ( 1 + A_1 x )^2 +
A_2 x^2 \right )^{3/2} + \ldots + A_k x^k \right )^{n_k} \;   ,
\ee
with $n_k$ defined in Eq. (17). The dependence on $t$ comes from the
quantities $A_n$ that are expressed through the coefficients $a_n(t)$ in
the process of the accuracy-through-order procedure. Then the large-variable
limit (45) yields
\be
\label{47}
A \left ( \left ( \ldots \left ( A_1^2 + A_2^2 \right )^{3/2} +
A_3 \right )^{4/3} + \ldots + A_k \right )^{(\bt-\al)/k} = B \; .
\ee
Since the right-hand sides in Eqs. (44), (45), or (47) do not depend on $t$,
the root approximant is {\it asymptotically} $t$-invariant.

Moreover, the root approximants, as has been shown, approximate well the
sought function in the whole interval of the variable $x \in [o, \infty)$.
More precisely, the root approximant $f^*_k(x,t)$ uniformly approximates the
sought function $f(x)$, with the maximal error $\varepsilon$, so that
\be
\label{48}
 \left | \; \frac{f_k^*(x,t) - f(x)}{f(x)} \; \right | < \ep \;  .
\ee
It is easy to see that if there are two approximants, related to two different
parameters $t_1$ and $t_2$, such that
$$
 \left | \; \frac{f_k^*(x,t_1) - f(x)}{f(x)} \; \right | < \ep_1 \; ,
\qquad
 \left | \; \frac{f_k^*(x,t_2) - f(x)}{f(x)} \; \right | < \ep_2 \; ,
$$
then the difference between these approximants is described by the inequality
\be
\label{49}
 \left | \; \frac{f_k^*(x,t_1) - f_k^*(x,t_2)}{f(x)} \; \right | <
\ep_1 + \ep_2 \;  .
\ee
In that sense, the root approximants, within the given accuracy, are
{\it approximately scheme invariant}, that is, invariant with respect to
the parameter $t$ labelling different perturbative schemes.

The dependence on a parameter can also occur in the change of the variables
$x = x(z,t)$, similarly to the scale dependence in field theory. Then it
is possible to expand $f_k(x(z,t))$ in powers of the new variable $z$,
getting $g_k(z,t)$. The corresponding root approximant $g^*_k(z,t)$, after
the inverse transformation $z = z(x,t)$, resulting in $g^*_k(z(x,t),t)$ is
such that, by construction, it uniformly approximates the sought function
$f(x)$, so that
\be
\label{50}
\left | \; \frac{g_k^*(z(x,t),t) - f(x)}{f(x)} \; \right | < \ep' \;   ,
\ee
within the maximal error $\varepsilon^{\prime}$. This allows us to classify
the root approximants as {\it approximately scale invariant}, within the
given accuracy.

We would like to note it again that the problem of interpolation is different
from that of extrapolation. For the latter, the problem of scheme and scale
invariance is more complicated. We plan to consider this in future
publications.

\section{Conclusion}

We have suggested a general and simple method for interpolation between
small-variable and large-variable asymptotic expansions. The method is based
on self-similar approximation theory, which allows for the construction of
approximations, whose form follows from extracting the properties of functional
similarity between the given expansion orders. Mathematical details of this
theory can be found in the cited references. The resulting self-similar root
approximant (6) makes it possible to satisfy the small as well as large
variable expansions and to uniformly describe the sought function in the whole
domain of its definition.

The general form of the root approximant (6) includes as particular cases that
of Pad\'{e} approximants, because of which the accuracy of root approximants
is not worse than that of Pad\'{e} approximants. But the root approximants enjoy
several advantages, as compared to Pad\'{e} approximants.

First, a root approximant for given orders $k$ and $p$ is uniquely defined,
while Pad\'{e} approximants for each given orders, allow for multiple
representations. Really, for a small-variable expansion of order $k$ and
large-variable expansion of order $p$, it is admissible to construct the whole
table of Pad\'{e} approximants $P_{M/N}$, with different $M$ and $N$ satisfying
the equality $M + N = k + p + 1$. Also, when one intends to construct diagonal 
Pad\'{e} approximants, one needs an even number of terms in a small-variable
expansion, while root approximants can be formed for any number of such terms. 

Second important advantage is that root approximants make it possible to
predict large-variable expansions, being based on small-variable expansions.
Such a prediction by means of Pad\'{e} approximants is principally impossible,
when the small-variable and large-variable expansions contain incompatible
powers. Moreover, in the case of incompatible expansions, Pad\'{e} approximants
cannot be defined at all. 

We have formulated uniqueness conditions allowing us to uniquely define all
parameters of the root approximant (6) from the coefficients of the given
asymptotic expansions. The root approximants are shown to be approximately
scheme and scale invariant.

The use of the interpolation formula (6) is convenient for the problems of
high energy physics, when, due to the duality between weak coupling and strong
coupling, there exist asymptotic expansions for both these limits. We have
illustrated our approach by several examples, demonstrating the generality,
simplicity, and good accuracy of this method.

\vskip 1cm

\section*{Acknowledgement}

One of the authors (V.I.Y.) is grateful to E.P. Yukalova for useful discussions.

\newpage

\begin{center}

{\bf {\Large Table Caption} }

\end{center}

\vskip 3cm

{\bf Table 1}: Ground-state energy of Schwinger model, for the varying
dimensionless parameter $x = m/g$, in different approximations: Density
matrix renormalization group, $E_{DMRG}$, Pad\'{e} approximants, $E_{PA}$,
fast moving frame estimates, $E_{FMFE}$, and self-similar root
approximant $E_5^*$.

\newpage

\begin{table}

\center

{\bf Table 1} \\ [5mm]

\begin{tabular}{|c|c|c|c|c|} \hline
$x$   & $E_{DMRG}$ & $E_{PA}$ & $E_{FMFE}$ & $E_5^*$  \\ \hline
0.125 & 0.540      & 0.540    & 0.528      & 0.540 \\  \hline
0.25  & 0.519      & 0.520    & 0.511      & 0.519 \\  \hline
0.5   & 0.487      & 0.489    & 0.489      & 0.487 \\  \hline
1     & 0.444      & 0.447    & 0.455      & 0.444 \\  \hline
2     & 0.398      & 0.396    & 0.394      & 0.392 \\  \hline
4     & 0.340      & 0.340    & 0.339      & 0.337 \\  \hline
8     & 0.287      & 0.286    & 0.285      & 0.284 \\  \hline
16    & 0.238      & 0.236    & 0.235      & 0.235 \\  \hline

\end{tabular}
\caption*{Ground-state energy of Schwinger model, for the varying dimensionless
parameter $x = m/g$, in different approximations: Density matrix
renormalization group, $E_{DMRG}$, Pad\'{e} approximants, $E_{PA}$,
fast moving frame estimates, $E_{FMFE}$, and self-similar root
approximant $E_5^*$. }

\end{table}

\end{document}